# $^{99,101}$Ru NMR and $^{63,65}$Cu NQR Study of the Magnetic Superconductors RuSr$_2$EuCu$_2$O$_8$ and RuSr$_2$GdCu$_2$O$_8$

Z. H. Han, J. I. Budnick, and W. A. Hines *
Department of Physics and Institute of Materials Science
University of Connecticut
Storrs, CT 06269-3046, USA

and

P. W. Klamut[1,2], M. Maxwell, and B. Dabrowski[1]
Department of Physics
Northern Illinois University
DeKalb, IL 60115, USA

ABSTRACT

In order to investigate the coexistence of magnetism and superconductivity on a microscopic scale, local studies using zero-field spin-echo $^{99,101}$Ru NMR and $^{63,65}$Cu NQR have been carried out on both superconducting and non-superconducting samples of magnetically-ordered RuSr$_2$EuCu$_2$O$_8$ and RuSr$_2$GdCu$_2$O$_8$. $^{99,101}$Ru NMR signals were observed at 1.3 K over two distinct frequency ranges. The high frequency spectra (110 to 150 MHz), which are essentially the same for all samples, are characterized by five sharp peaks for each isotope corresponding to a hyperfine field of approximately 590 kOe with the appropriate quadrupole splitting. The low frequency spectra (40 to 90 MHz), which show significant variation from sample to sample, correspond to a much lower hyperfine field with considerable broadening and quadrupole features that are barely resolvable. As reported previously, the low frequency spectra are attributed to the Ru$^{4+}$ valence state while the high frequency spectra are attributed to the Ru$^{5+}$ valence state. $^{63,65}$Cu NQR features were observed (26 to 34 MHz), although with considerable broadening. The results are discussed in terms of the microscopic magnetic structure, mixed valence state for Ru, occupancy of the Ru sites, and the existence and role of impurity phases.

[1] also in the Materials Science Division, Argonne National Laboratory, Argonne, IL 60439, USA
[2] also in the Institute of Low Temperatures and Structure Research of the Polish Academy of Sciences, 50-950 Wrocław, Poland
* Corresponding author. *Email address*: hines@uconnvm.uconn.edu (W.A. Hines).

I. INTRODUCTION

The possible coexistence of superconductivity and magnetic ordering (in particular, ferromagnetism) on a microscopic scale has been investigated and discussed for many years. For several homogenous intermetallic compounds, despite the orbital pair breaking effect which would prevent singlet-state pairing within a ferromagnetic state, the two states were found to coexist over a narrow temperature range with spatially-modified magnetic and superconducting order parameters. This effect, the notable examples being $ErRh_4B_4$ [1,2] and $HoMo_6S_8$ [3], hints at a nanometer-range domain structure developed to accommodate both phases [2]. Also, later reports of a few other intermetallic ferromagnetic superconductors, in which triplet state pairing coexists with a comparatively weak long-range ferromagnetic ordering, has refueled interest in this subject [4]. By investigating the details of simultaneous accommodation of the two states, one might expect to also shed light on the nature of the superconducting state itself.

Recently, the synthesis of a family of ruthenocuprate materials has contributed to this debate with compounds which belong to the celebrated high-$T_c$ superconductor family [5,6]. The first report of the simultaneous observation of both superconductivity and magnetic ordering for a ruthenocuprate compound was $RuSr_2RE_{1.4}Ce_{0.6}Cu_2O_{10-\delta}$ (where RE = rare earth, Eu or Gd), commonly referred to as a "Ru1222" type [7]. Not long after this report, another ruthenocuprate compound, $RuSr_2GdCu_2O_8$ or "Ru1212" type, was found to have a magnetic ordering transition $T_o$ at 133 K and a bulk Meissner state for temperature $T_c$ less than 40 K [8,9]. This was followed by the discovery of similar behavior in the Ru1212 compounds $RuSr_2EuCu_2O_8$ ($T_o$ = 132 K, $T_c \leq$ 25 K) [10-12], $RuSr_2YCu_2O_8$ ($T_o$ = 149 K, $T_c \leq$ 39 K) [13,14]. $RuSr_2RECu_2O_8$ compounds also have been reported for RE = Dy, Ho, and Er [15]. For all of the Ru1212 materials, the temperature values for onset of superconductivity as well as for zero resistance, are critically dependent on the sample preparation conditions. The crystal structure for $RuSr_2RECu_2O_8$ is similar to that for $YBa_2Cu_3O_7$ except that the one-dimensional Cu-O chains are replaced by two-dimensional $RuO_2$ square planar layers. Thus, from the point of view of superconductivity, the role of the $RuO_2$ planes, which are mainly responsible for the magnetism, would be to act as the charge reservoir which is necessary to dope holes into the $CuO_2$ planes. The crystal symmetry is essentially tetragonal; however, there is an additional complexity when compared to $YBa_2Cu_3O_7$. High-resolution electron microscopy and synchrotron x-ray diffraction studies on $RuSr_2GdCu_2O_8$ have shown that the $RuO_6$ octahedra are coherently rotated about the c-axis within subdomains of 5 to 20 nm [16]. In addition, a slight tilting of the $RuO_6$ was also observed [16].

Although numerous measurements of the physical properties for both the Ru1212 and Ru1222 materials have been carried out, significant questions still remain concerning the microscopic magnetic structure, the mixed valence state for Ru, occupancy of the Ru sites, and also the existence and role of impurity phases. Zero-field muon spin rotation measurements on $RuSr_2GdCu_2O_8$ indicate that the bulk magnetism is due to an ordering of the Ru moments; it is homogenous on a microscopic scale and accounts for most of the sample volume [9]. It has also been suggested that the magnetic order is not significantly modified by the onset of superconductivity [9]. However, the exact nature of the magnetic ordering of the Ru moments is still in doubt. In particular, powder neutron diffraction measurements for both $RuSr_2GdCu_2O_8$ and $RuSr_2EuCu_2O_8$ show that the magnetic ordering has the G-type antiferromagnetic structure in which the neighboring Ru moments are antiparallel in all three directions, with a low temperature value for the ordered moment $\mu(Ru) \approx 1.2$ $\mu_B$ along the c-axis [13,17-19]. On the other hand, measurements of the dc magnetization indicate weak ferromagnetic behavior due to the existence of hysteresis loops with remanence values about 15% of the saturation values and coercive fields as high as 400 Oe. Low-temperature and high-field magnetization measurements yield $\mu_{sat}(Ru) \leq 1.0$ $\mu_B$. Curie-Weiss fits to the high-temperature magnetic susceptibility yield positive $\Theta$ values, indicating ferromagnetic interactions between the Ru moments, and $\mu_{eff}(Ru) \approx 3.2$ $\mu_B$ [12,20]. However, it should be noted that cusp-like behavior in the low-field magnetic susceptibility exists near the magnetic ordering temperature which is characteristic of antiferromagnetic ordering. One possible scenario that has been argued to account for the observation of both ferromagnetic and antiferromagnetic behavior is a canting of antiferromagnetically ordered Ru-moments. An alternative explanation has emerged from recent x-ray absorption near edge structure (XANES) measurements on $RuSr_2GdCu_2O_8$ which indicate the existence of a mixed valence state with 40-50% $Ru^{4+}$ and 60-50% $Ru^{5+}$ [21]. The existence of a mixed valence state with approximately equal amounts of $Ru^{4+}$ and $Ru^{5+}$ is also consistent with other nuclear magnetic resonance (NMR) results [22-25] as well as the NMR results presented here. Assuming a mixed Ru valence, along with the corresponding two quite different localized moments within a $RuO_2$ plane, allows for the possibility of ferrimagnetic ordering such that adjacent moments are antiparallel; however, they sum to a net moment because of their two different magnitudes. A model such as this could explain the different values of the high-temperature effective moment ($\mu_{eff}(Ru) \approx 3.2$ $\mu_B$) and the low-temperature "saturation" moment ($\mu_{sat}(Ru) \leq 1.0$ $\mu_B$). In addition, from their mixed valence results Liu et al. [21] estimate values $p \approx 0.2$ for the doped hole concentration in the $CuO_2$ planes which are larger than the values obtained from transport measurements ($p \approx 0.07$) [26], but smaller than the values obtained from measurements of the Cu-O bond distances ($p \approx 0.4$) [16].

A major debate is still ongoing concerning the intrinsic properties of the Ru1212 phase and possible modifications caused by the occurrence of intersite substitutions and/or cation defects present within the host matrix. The importance of the sample synthesis procedure in any attempt to obtain a homogeneous single phase material must be stressed. Both superconducting and non-superconducting samples of $RuSr_2GdCu_2O_8$ have been reported, with nominally the same stoichiometry, which raises questions concerning the details of the superconducting phase and sample homogeneity [5-7,27,28]. Recent x-ray synchrotron and neutron diffraction experiments by Blake et al. [29] revealed a slight difference in the structural parameters between superconducting (SC) and non-superconducting (NSC) samples (SC: c/a = 3.0162, NSC: c/a = 3.0145). Their structural analysis also allows for partial Cu-Ru substitutions to be present within the Ru1212 phase. This should be noted in view of increased superconducting $T_c$ as well as strongly suppressed magnetism found in the $Ru_{1-x}Sr_2GdCu_{2+x}O_8$ phases for $x > 0$ [28,30]. A slight decrease of the magnetic transition temperature has also been observed for superconducting relative to non-superconducting samples of $RuSr_2GdCu_2O_8$ [28]. When discussing phase homogeneity, the presence of minor impurity phases (at a level of a few volume percent) should be also addressed. For example, a review of the literature clearly shows the difficulty of eliminating $SrRuO_3$, or perhaps the substituted $Sr(Gd)Ru(Cu)O_3$, as a second phase in these materials. As a consequence, there is the possibility that either the magnetic or superconducting characteristics might be related to a minority second phase as well. The muon spin rotation experiments cited above provide strong evidence that the magnetic phase persists throughout the entire volume of the sample [9]. Concerning the homogeneity of the superconducting phase, Bernhard et al. [31] provide as evidence for a bulk Meissner state in $RuSr_2GdCu_2O_8$: (1) the sizeable diamagnetic response at low temperature in the dc magnetization and (2) the sizable peak at $T_c$ in the heat capacity measurements (see also ref. [26]). Arguments for a spontaneous vortex phase in the intermediate temperature range below $T_c$ were also presented. These authors also report that both high-resolution synchrotron x-ray diffraction and neutron diffraction measurements on their Ru1212 samples indicate a high structural and chemical homogeneity with no detectable impurity phases above the limits of sensitivity (~ 1%) [31]. On the other hand, Awana et al. [32] argue that Ru and Cu cannot be distinguished without ambiguity by neutron diffraction and, furthermore, that Ru1212 samples may not be homogeneous in composition. They claim that Ru/Cu ordering at the charge-reservoir cation site is a likely possibility. Based on their transmission electron microscope diffraction results for $RuSr_2GdCu_2O_8$, they suggest that there are regions of a superconducting $Ru_{0.5}Cu_{0.5}Sr_2GdCu_2O_8$ minority phase in $RuSr_2GdCu_2O_8$, and refute the coexistence of superconductivity and magnetism in intrinsically pure Ru1212 [32]. Yokosawa et al. [33] recently reported electron microscopy studies which suggest the existence of superlattice domains with period of approximately 10 nm for both Ru1212 and Ru1222 materials. The existence of domains was attributed to reversed rotations of $RuO_6$ octahedra about the c-axis. A description in which domains separated by antiphase boundaries with a reversed sense of rotation of $RuO_6$ octhaedra has also been suggested in earlier work [16,18]. We note that the possible alternation of magnetic structure at domain boundaries could lead to the interesting situation of having both AFM and FM components in the Ru spin sublattice.

An analysis of the stability of the Ru1212 phase presented by Zhigadlo et al. [34] revealed that partial decomposition occurs at 1050°C, i.e., a slightly lower temperature than 1060°C which is commonly reported for synthesis of the superconducting material. The resulting high temperature phases were found to recombine with fast kinetics upon cooling through the 1050°C threshold. Such a reaction path could promote stabilization of the microsize variations in the crystal structure [34]. In accordance with the properties of the $Ru_{1-x}Sr_2GdCu_{2+x}O_{8-y}$ phases, if structural regions with Ru partially substituted by the Cu atoms are formed in the Ru1212 matrix, these regions would be superconducting with $T_c$ defined by locally modified oxygen content [28,30]. Such structural variations also could be responsible for the formation of a superstructure. In the early work by Tallon et al. [8], the superlattice observed in electron diffraction patterns was thought to originate in the ordered intersubstitutions of cations. Phase separation issues were discussed in detail by Xue et al. [35] and Lorenz et al. [36], where arguments were given for the occurrence of spatial separation between ferromagnetic and antiferromagnetic regions in the sample, with superconductivity occurring in the latter. Microscopic phase separation, driven by the tendency of an electronic system to minimize its energy, remains a tempting description for the coexistence of superconductivity and ferromagnetism. For ruthenocuprates, however, the possibility of developed structural inhomogeneity should also be taken into account.

In order to gain insight into the coexistence of magnetism and superconductivity on a microscopic scale in the Ru1212 phases, local studies using zero-field spin-echo $^{99,101}$Ru NMR and $^{63,65}$Cu NQR have been carried out on magnetically-ordered $RuSr_2EuCu_2O_8$ and $RuSr_2GdCu_2O_8$. The results presented and discussed here are for both superconducting and non-superconducting samples of these materials.

## II. EXPERIMENTAL APPARATUS AND PROCEDURE

Batches of polycrystalline $RuSr_2EuCu_2O_8$ and $RuSr_2GdCu_2O_8$ materials were synthesized by a solid state reaction technique that involved stoichiometric oxides of $RuO_2$, either $Eu_2O_3$ or $Gd_2O_3$, CuO, and $SrCO_3$. The solid state reaction consisted of calcination in air at 920 °C for 12 hours. The materials were then ground, pressed into pellets, and subjected to various heat treatments as follows. For batch (a), hereafter referred to as "non-superconducting $RuSr_2EuCu_2O_8$", the material was annealed at 930-935 °C in the flow of 1% oxygen in argon. There were several such annealings, with an intermediate grinding and pelletizing, performed in this atmosphere and they were always followed by cooling in argon to avoid the reformation of the $SrRuO_3$–type impurity phase. This method of synthesis has been shown previously to result in single-phase, non-superconducting material for both $RuSr_2EuCu_2O_8$ and $RuSr_2GdCu_2O_8$ [37]. For batch (c), hereafter referred to as "non-superconducting $RuSr_2GdCu_2O_8$", the material was prepared using the same synthesis procedure as described above for batch (a). For batch (b), hereafter referred to as "partially-superconducting $RuSr_2EuCu_2O_8$", the material was also annealed at 930 °C in the flow of 1% oxygen in argon as for (a) and (c). The material was then annealed twice in flowing oxygen at 1060-1065 °C for 90 hours, followed by slow cooling. Finally, for batch (d), hereafter referred to as "superconducting $RuSr_2GdCu_2O_8$", the material was annealed at 970 °C in flowing oxygen. The material was then sintered at 1060 °C for 10 hours in a high-pressure oxygen atmosphere (600 bar). This was followed by annealing in flowing oxygen at 1060 °C for seven days, followed by a slow cooling. Incorporation of the high pressure oxygen annealing step, which also stabilizes the x>0 phases of $Ru_{1-x}Sr_2GdCu_{2+x}O_{8-y}$ [30], improved the superconducting characteristics of this sample [38]. All four batches of material were characterized by using x-ray diffraction (XRD) with a Bruker powder diffractometer and Cu $K_{\alpha I}$ radiation ($\lambda$ = 1.5406 Å). In addition, magnetization measurements were carried out on a Quantum Design SQUID magnetometer; temperature dependencies of the *ac* susceptibility were measured with a Quantum Design susceptometer (PPMS system).

Samples suitable for NMR/NQR were prepared by placing approximately 60 mg of powder in solenoidal coils (2.0 mm diameter, 4.0 mm length) wound for the frequency ranges $26 \leq \nu \leq 34$, $40 \leq \nu \leq 90$, and $110 \leq \nu \leq 150$. Zero-field spin-echo NMR (NQR) signals were obtained at 1.3 K from the $^{99,101}$Ru ($^{63,65}$Cu) nuclei over the frequency ranges above using a modified Matec model 5100 mainframe and model 525 gated RF amplifier in combination with a model 625 broadband receiver with phase coherent detection. The NMR signals were optimized using a standard spin-echo pulse sequence by adjusting the power for RF pulses ranging from 1.0 to 2.0 µs in duration. Typically, the pulse separation was 20 µs and the pulse sequence repetition rate approximately 20 Hz. Useable spectra were obtained by averaging the NMR/NQR signals 500 to 1000 times at appropriately spaced frequency intervals across the spectrum. In addition, measurements of the spin-spin relaxation time $T_2$ were made at selected frequencies by varying the separation $\tau$ between the two RF pulses. A temperature of 1.3 K was obtained by pumping on liquid He in a conventional double dewar system.

The NMR/NQR spectrometer described above is a 50 Ω system. In these experiments, the probe coax, which extended down into the liquid helium, was terminated by the sample coil in series with a variable capacitor which could be tuned from the top of the dewar. The matching was achieved using a fixed inductor, also at the end of the coax, which was in parallel with the series tuned circuit. This scheme has the advantage that a single adjustment enables one to tune over a relatively wide range of frequency, while maintaining a satisfactory match to 50 Ω [39]. A second coax, which also extended down into the liquid helium, was terminated by a 50 Ω resistor near the sample coil. This served as a broadband antenna which was used to inject a calibration signal as well as monitor the pulsed RF field. The reader is referred to Zhang et al. [40] and references therein for additional details concerning zero-field spin-echo NMR.

## III. EXPERIMENTAL RESULTS AND ANALYSIS

Figure 1 shows the XRD scans obtained for the four samples studied in this work (non-superconducting and superconducting samples for both $RuSr_2EuCu_2O_8$ and $RuSr_2GdCu_2O_8$). All four samples were indexed using a tetragonal unit cell (P4/mmm) and yielded lattice parameters consistent with those reported previously [5]. The XRD scan for the partially-superconducting $RuSr_2EuCu_2O_8$ sample (Fig. 1b) shows the presence of the $SrRuO_3$ impurity phase; the impurity phase is not seen in the scans for the other samples. Also, the XRD scan for the superconducting $RuSr_2GdCu_2O_8$ sample (Fig. 1d) shows that the (200) and (006) reflections for 2θ near 47.5° are clearly resolved. This observation is a favorable indicator for the occurrence of superconductivity in high-Tc cuprate systems with perovskite-based structures [41]. These two reflections are not resolved for the other three samples.

Figure 2 shows the zero-field-cooled (ZFC) and field-cooled (FC) magnetization as a function of the temperature for the indicated magnetic field. All four samples show the onset of magnetic ordering for temperatures ranging between 130 K and 145 K. The magnetic ordering temperatures for the samples, which were obtained by taking the derivative of the FC magnetization curve, are listed in Table I. In addition, the superconducting $RuSr_2GdCu_2O_8$ sample(Fig. 2d) shows a superconducting transition with an onset temperature at approximately 30 K as mapped in the *dc* magnetization. Figure 3 presents the temperature dependencies of the *ac* susceptibility and detail of the superconducting transitions for the samples (b) and (d). Presented there $T_{c1}$ and $T_{c2}$ temperatures have been frequently used in the literature for characterizing the superconducting transition in Ru1212 compounds. The $T_{c1}$ mark the onset of intrinsic superconducting transition, the $T_{c2}$ - the onset for establishing bulk screening currents, which also corresponds to a sharp increase of the lossy component (not shown) of the *ac* susceptibility (see also refs. 28 and 38 for more discussion of these characteristics). One should note, that the small peak in the ZFC magnetization curve and plateau in the FC magnetization curve upon entering the superconducting state (Fig.2d) are the features which have not been observed in all reported studies of the superconducting $RuSr_2GdCu_2O_8$. Although similar peak-like behavior is observed for about the same temperature range in the magnetic behavior of the $Sr_2GdRuO_6$ impurity phase (candidate for the trace impurity phase), Papageorgiou et al. [42] have excluded this as a possible explanation for the features observed for $RuSr_2GdCu_2O_8$. Instead, they show that these features are related to the superconductivity of $RuSr_2GdCu_2O_8$ (see also [38]) and due to a very low screening capability of this material in a vicinity of its intrinsic superconducting transition. The anomaly could then be caused by motion of the weakly pinned flux, generated by the movement of the sample in somewhat inhomogeneous magnetic field of the SQUID during measurement [42]. Finally, the partially-superconducting $RuSr_2EuCu_2O_8$ sample (Figs. 2b, 3) shows a weak superconducting onset temperature at 28 K; however, the shielding is far from complete even at the lowest accessed temperature. The onset temperatures are also listed in Table I.

In order to obtain information about the magnetic moment associated with Ru, additional magnetization measurements were carried out on the non-superconducting and partially-superconducting $RuSr_2EuCu_2O_8$ samples in both the magnetically-ordered and paramagnetic states. Unlike the case for Gd, $Eu^{3+}$ ions carry essentially no moment and the analysis is reasonably straightforward. Figures 4a and 4b show the full hysteresis loops measured at 5.0 K for non-superconducting and partially-superconducting $RuSr_2EuCu_2O_8$, respectively. Although the magnetization does not saturate in either case, Ru moment values of 0.99 $\mu_B$ and 0.95 $\mu_B$ were obtained for non-superconducting and partially-superconducting $RuSr_2EuCu_2O_8$, respectively, in the magnetically-ordered state. The above values are calculated from the measured magnetization for 5.0 K and 50 kOe. Figures 5a and 5b show the Curie-Weiss analysis for non-superconducting and partially-superconducting $RuSr_2EuCu_2O_8$, respectively, above the ordering temperature in the paramagnetic state over the temperature range 205 K $\leq$ T $\leq$ 305 K. For temperatures within the above range, the measured magnetization versus magnetic field was completely linear through the origin yielding values for the paramagnetic susceptibility. Fits of the paramagnetic susceptibility were made to the Curie-Weiss law

$$\chi = \frac{N\mu_{eff}^2}{3k_B(T-\Theta)} + \chi_0 \qquad \text{eqn. (1)}$$

where N is the concentration of Ru moments, $\mu_{eff}$ is the effective Ru moment (magnitude), $k_B$ is the Boltzmann constant, $\Theta$ is the Curie-Weiss temperature, and $\chi_0$ is a temperature independent term which reflects the core diamagnetism, Landau diamagnetism, and Pauli paramagnetism. The parameters obtained from the Curie-Weiss fits are listed below in Table II. Ru moment (magnitude) values of $3.2_0$ $\mu_B$ and $3.1_9$ $\mu_B$ were obtained for non-superconducting and partially-superconducting $RuSr_2EuCu_2O_8$, respectively, in the paramagnetic state. These values fall in between the values expected for $Ru^{4+}$ (1.83 $\mu_B$) and $Ru^{5+}$ (3.87 $\mu_B$), suggesting a mixed valence state for the Ru atoms.

The NMR/NQR spectra for the four samples, which were obtained at 1.3 K, are shown in Figure 6. In the analysis of the spectra, three distinct frequency ranges can be considered (26 $\leq \nu \leq$ 34, 40 $\leq \nu \leq$ 90, and 110 $\leq \nu \leq$ 150). The high frequency spectra (110 to 150 MHz), which are essentially the same for all samples, are attributed to zero-field NMR signals from the $^{99}$Ru ($\gamma$ = 0.19645 MHz/kOe, I = 5/2, and 12.7% abundance) and $^{101}$Ru ($\gamma$ = 0.22018 MHz/kOe, I = 5/2, and 17.1% abundance) isotopes [43]. As reported previously by Kumagai et al. [22] for $RuSr_2GdCu_2O_8$, the $^{99,101}$Ru NMR signals which occur in this frequency range arise from Ru atoms in sites characterized by the $Ru^{5+}$ valence state with S = 3/2 [22]. There are five sharp peaks for each isotope corresponding to an internal hyperfine field $H_{hf}$ of approximately -590 kOe with the appropriate quadrupole splitting, described by $\nu_Q$. In particular, the NMR peak frequencies can be described by assuming a dominant magnetic Zeeman interaction along with a quadrupole perturbation which must be calculated to second order where the asymmetry factor $\eta \approx 0$ due to the tetragonal structure. All of the spectra were fit by taking an angle between the direction of the internal field and the principal axis of the electric field gradient $\theta \approx 90°$, along with the appropriate values of $H_{hf}$ and $\nu_Q$. The negative sign for $H_{hf}$ is confirmed by the application of an external magnetic field. The particular values for $H_{hf}$ and $\nu_Q$ which results from the fits to the spectra are listed in Table III. The values in Table III are completely consistent with other work reported during these studies [22-25]. Also, there were no detectable NMR signals at 119.6 MHz and 133.1 MHz for the four samples studied in this work. These frequencies characterize the $^{99,101}$Ru peaks, respectively, for the $Sr_2GdRuO_6$ candidate impurity phase in which all of the Ru atoms are in the $Ru^{5+}$ valence state. $^{99,101}$Ru NMR zero-field signals are also observed over a lower frequency range (40 to 90 MHz). Unlike the high frequency spectra described above, these spectra, which show significant variation from sample to sample, correspond to a much lower hyperfine field with considerable broadening and quadrupole features that are barely resolvable. The low frequency spectra are attributed to the $Ru^{4+}$ valence state with S = 1 [22]. Also, unlike the high frequency spectra, good quality fits were not possible and only estimates for $H_{hf}$ were obtained (see Table III). Finally, $^{63,65}$Cu zero-field NQR features were observed (26 to 34 MHz), although with considerable broadening. Only for the superconducting $RuSr_2GdCu_2O_8$ spectrum are two distinct peaks resolved (see Fig. 4d). The peaks at 28.2 MHz and 30.4 MHz (frequency ratio = 1.08) correspond to the $^{65}$Cu and $^{63}$Cu nuclei, respectively (quadrupole moment ratio = 1.08).

The behavior of the $Ru^{5+}$ spectra in the presence of an external magnetic field was investigated. Figure 7 shows the evolution of the zero-field spin-echo $^{101}$Ru NMR central (-1/2 ↔ +1/2) peak at 1.3 K for the superconducting $RuSr_2GdCu_2O_8$ sample. Similar behavior was observed for the partially-superconducting $RuSr_2EuCu_2O_8$ sample. From spectra such as that shown in Fig. 7, the central peak height (amplitude maximum) as a function of the applied magnetic field was determined for the partially-superconducting $RuSr_2EuCu_2O_8$ (Fig. 8a) and superconducting $RuSr_2GdCu_2O_8$ (Fig. 8b) samples. For both cases, the peak height increases, reaching a maximum between 4 to 6 kOe, and then decreases. This behavior is not typical for a bulk multidomain ferromagnetic material. In a typical ferromagnetic material, the NMR signals are dominated by nuclei in the domain walls due to the enhancement factor. Consequently, the application of an external magnetic field progressively eliminates the domain walls and reduces the NMR signal intensity. Also, from spectra such as that shown in Fig. 7, the central peak position as a function of the applied magnetic field was determined for the partially-superconducting $RuSr_2EuCu_2O_8$ (Fig. 9a) and superconducting $RuSr_2GdCu_2O_8$ (Fig. 9b) samples. For both cases, the peak position shifts to lower frequency with a linear behavior for higher fields. This behavior is not typical for a bulk polycrystalline antiferromagnetic material. For a typical polycrystalline antiferromagnetic material, the application of an external magnetic field would result in a broadening of the NMR peak, but essentially no change in position. It is important to note that the application of an external magnetic field resulted in a systematic shift of the entire five peak $Ru^{5+}$ spectrum to lower frequency with essentially no change in the quadrupole splitting (see Fig. 7 inset). Finally, the behavior of the $Ru^{4+}$ spectra in the presence of an externally applied magnetic field is the same as that described above for the $Ru^{5+}$ spectra. On the basis of the field dependence described above, we believe that the magnetic state is more complicated than previously suggested.

IV. DISCUSSION AND CONCLUSIONS

As described above, $^{99,101}$Ru NMR signals were observed over two distinct frequency ranges. It is noteworthy that NMR work on the Ru1212 systems indicates the existence of two Ru valence states, namely $Ru^{4+}$ and $Ru^{5+}$, although there is only one formal crystallographic site in the Ru1212 structure. The high frequency spectra (110 to 150 MHz), which are essentially the same for all samples, are attributed to Ru atoms in having the $Ru^{5+}$ valence state. The low frequency spectra (40 to 90 MHz), which show significant variation from sample to sample, correspond to a much lower hyperfine field with considerable broadening and quadrupole features that are barely distinguishable (indicating more disorder). The low frequency spectra are attributed to Ru atoms in the $Ru^{4+}$ valence state. Based on the integrated intensities over the two spectral ranges, it is reasonable to conclude that the two types of Ru atoms are approximately equal. The conclusion that Ru is in a mixed valence state is supported by recent Ru $L_{III}$ x-ray absorption near-edge spectroscopy (XANES) work where a mixed valence of 40-50% $Ru^{4+}$ and 60-50% $Ru^{5+}$ is reported [21]. From this result, a Cu valence is determined to be $\approx$ 2.20-2.25, corresponding to a hole concentration $\approx$ 0.20-0.25, which is comparable to that for $YBa_2Cu_3O_{7-\delta}$. On the other hand, the results from recent Mössbauer experiments are puzzling [44,45]. Measurements of the isomer shift by Kruk et al. [44] below the ordering temperature yield an intermediate value which is consistent with a mixed valence state; however, DeMarco et al. [45] report that only one value for the $H_{hf}$ and isomer shift (which is attributed to $Ru^{5+}$) is obtained from fits to the spectra below the ordering temperature. It is of interest to realize that the $Ru^{5+}$ NMR spectrum remains very well resolved for both the $^{99}$Ru and $^{101}$Ru isotopes, suggesting that a well ordered local structure determines the $Ru^{5+}$ hyperfine interaction. The $Ru^{4+}$ signal is extremely broad in all samples except for the superconducting $RuSr_2GdCu_2O_8$ sample in which the relatively narrower and better resolved $Ru^{4+}$ signal permits the observation of the $^{63,65}$Cu NQR signal.

As described above in Section III, measurements of the paramagnetic susceptibility above the ordering temperature support the idea of a mixed valence state in the Ru1212 systems. In particular, the Curie-Weiss fits which were carried out for the $RuSr_2EuCu_2O_8$ samples (both non-superconducting and partially-superconducting) yielded effective moment (magnitude) values near 3.2 $\mu_B$ for Ru. This value is intermediate between that calculated for $Ru^{4+}$ (where S = 1 yields $\mu_{eff}$ = 1.83 $\mu_B$) and $Ru^{5+}$ (where S = 3/2 yields $\mu_{eff}$ = 3.87 $\mu_B$), where four electrons and three electrons, respectively, are assumed to occupy the crystal-field-split $t_{2g}$ states. It is possible that the difference in the electronic structure associated with the two types of Ru sites is related to the large disorder observed for the Ru1212 $Ru^{4+}$ NMR spectra, in contrast to the very sharp local environment responsible for the well-resolved $Ru^{5+}$ spectra. However, measurements of the Ru moment (component) below the ordering temperature yielded smaller values of 0.99 $\mu_B$ and 0.95 $\mu_B$, respectively, for the same two $RuSr_2EuCu_2O_8$ samples. The general trend of reduced $Ru^{4+}$ ordered moment in Ru1212 samples could be interpreted as demonstrating a certain degree of itinerancy of the Ru 4d electrons. Additional information concerning the Ru moment can be obtained from the NMR measurements of $H_{hf}$ by assuming that core polarization of the 4d electrons is the dominant contribution to $H_{hf}$ and taking a typical value for the hyperfine coupling constant to be $\approx$ -300 kOe/$\mu_B$ [46]. For $H_{hf} \approx$ -590 kOe, which characterizes the $Ru^{5+}$ spectra for all of the samples, $\mu(Ru^{5+}) \approx$ 2.0 $\mu_B$. This value is in good agreement with neutron powder diffraction measurements on the type I antiferromagnetic insulator $Sr_2YRuO_6$ in which only the $Ru^{5+}$ valence state exists and a value of 1.85 $\mu_B$/Ru is reported [47]. On the other hand, $H_{hf}$ ranges from -270 to -370 kOe for the $Ru^{4+}$ spectra obtained from the four Ru1212 samples, which yields $\mu(Ru^{4+}) \approx$ 0.90 to 1.2 $\mu_B$. For comparison, early neutron powder diffraction work on $SrRuO_3$, an itinerant ferromagnet in which only the $Ru^{4+}$ exists, yielded an ordered moment of 1.2 $\mu_B$ [48]. Using $H_{hf} \approx$ -329 kOe obtained from $^{99,101}$Ru NMR on $SrRuO_3$, along with the hyperfine coupling constant above, results in $\mu(Ru^{4+}) \approx$ 1.1 $\mu_B$ [49]. Magnetization measurements of the (non-saturated) ordered moment for $SrRuO_3$ range from 0.8 to 1.6 $\mu_B$/Ru [50]. The mixed state moment values for Ru in the Ru1212 systems which are obtained from NMR experiments do not appear to be consistent with either powder neutron diffraction measurements or measurements of the "saturation" magnetization. As referred to above in Section I, powder neutron diffraction measurements in these systems indicate a G-type antiferromagnetic ordering of the Ru moments along the c-axis which were fitted to a single low temperature value of μ(Ru) ≈ 1.2 $\mu_B$ [13,17-19], while magnetization measurements yield $\mu_{sat}$ ≤ 1.0 $\mu_B$ (see above).

As described above in Section III, the $Ru^{5+}$ NMR spectra, which are essentially the same for all four samples, are characterized by five sharp peaks for each isotope. Good fits to the spectra were obtained by taking a dominant magnetic Zeeman interaction along with a second order quadrupole perturbation in which the asymmetry factor η ≈ 0 due to the tetragonal crystal structure. In addition to values for $H_{hf}$ and $\nu_Q$, all of the spectral fits required θ ≈ 90°, for the angle between the direction of the internal field (which is responsible for the hyperfine interaction of the $Ru^{5+}$ NMR) and the principal axis of the electric field gradient tensor. This result is also inconsistent with the above-mentioned neutron diffraction work [13,17-19]. A resolution of the different magnetic structures resulting from the neutron diffraction experiments and from NMR work, including that reported here, may be found in studies of the domain and domain wall structure of the Ru1212 system. It is conceivable that the dominant enhanced $Ru^{5+}$ NMR signal, which is used to deduce the magnetization direction, may arise from very localized, easily driven magnetization within a domain wall. This would yield a magnetization direction which is different from that seen in the domains by the neutron diffraction experiments.

Thus, a puzzle remains on how to relate the magnetic structure derived from the neutron diffraction results and that deduced from the $Ru^{5+}$ signal. The neutron diffraction analysis is based on a single moment in this system, even though the results presented here suggest that both $Ru^{4+}$ and $Ru^{5+}$ are both present in nearly equal amounts and are strongly coupled in some, as not yet determined, local magnetic structure. A possible model which can be deduced from the magnetic field dependence of the $Ru^{5+}$ NMR spectrum features a simple antiferromagnetic order with the type-I structure, i.e., the Ru moments within a given $RuO_2$ are aligned ferromagnetically with an antiferromagnetic alignment between adjacent $RuO_2$ planes. However, such a picture cannot be argued conclusively.

In addition to the $Ru^{4+}$ and $Ru^{5+}$ NMR spectra which were observed over the frequency ranges 40 to 90 MHz and 110 to 150 MHz, respectively, $^{63,65}Cu$ NQR features were observed from 26 to 34 MHz, although with considerable broadening. It is noteworthy to compare the excitation conditions for the Ru and Cu signals as they are significantly different. For both the $Ru^{4+}$ and $Ru^{5+}$ NMR spectra, relatively low power was required due to the NMR enhancement factor. The NMR enhancement factor arises from the coherent motion of the magnetic moments with the RF pulses, indicating a ferromagnetic nature along with small magnetic anisotropy. On the other hand, relatively large power was needed to excite the Cu signals. This is consistent with a magnetic structure in which the local magnetic order exists primarily in the $RuO_2$ planes and does not extend into the $CuO_2$ planes. As a result there is no significant enhancement of the RF excitation field at the Cu nuclei.

Finally, the NMR results reported here imply a strong equal participation of $Ru^{4+}$ and $Ru^{5+}$ in the Ru ordering. From the behavior of the $Ru^{4+}$ and $Ru^{5+}$ signals in an externally applied field, we conclude that the two sites are intrinsic to the Ru1212 phase, and not due to any major presence of $SrRuO_3$ as an impurity phase. The $^{99,101}Ru$ NMR signals from $SrRuO_3$ would behave as either;

(1) a typical multidomain ferromagnetic material, in which there is a monotonic decrease in amplitude with field or (2) a nanoscale single domain ferromagnetic material, in which the amplitude is first constant and then decreases with field. Further work to directly explore the magnetic structure of the Ru moments is needed.

ACKNOWLEDGMENTS

The work at the University of Connecticut was supported by NSF grant DMR-9705136. The work at Northern Illinois University was supported by NSF grant DMR-0105398 and the State of Illinois under HECA. The authors wish to thank B. O. Wells, D. M. Pease, and M. Aindow for several useful discussions.

REFERENCES

[1]   W. A. Fertig, D. C. Johnston, L. E. DeLong, R. W. McCallum, M. B. Maple, and
      B. T. Matthias, Phys. Rev. Lett. 38, 987 (1977).
[2]   S. K. Sinha, G. W. Crabtree, D. G. Hinks, and H. Mook, Phys. Rev. Lett. 48, 950 (1982).
[3]   M. Ishikawa and Ø. Fischer, Solid State Commun. 23, 37 (1977).
[4]   For a recent review of this topic, see J. Flouquet and A. Buzdin,
      http://physicsweb.org/article/world/15/1/9
[5]   L. Bauernfeind, W. Widder, and H. F. Braun, Physica C 254, 151 (1995).
[6]   L. Bauernfeind, W. Widder, and H. F. Braun, J. Low Temp. Phys. 105, 1605 (1996).
[7]   I. Felner, U. Asaf, Y. Levi, and O. Millo, Phys. Rev. B 55, R3374 (1997).
[8]   J. Tallon, C. Bernhard, M. Bowden, P. Gilberd, T. Stoto, and D. Pringle, IEEE Trans.
      Appl. Supercond. 9, 1696 (1999).
[9]   C. Bernhard, J. L. Tallon, Ch. Niedermayer, Th. Blasius, A. Golnik, E. Brücher,
      R. K. Kremer, D. R. Noakes, C. E. Stronach, and E. J. Ansaldo, Phys. Rev. B 59, 14099
      (1999).
[10]  K. Otzschi, T. Mizukami, T. Hinouchi, J. Shimoyama, and K. Kishio, J. Low Temp. Phys.
      117, 855 (1999).
[11]  R. L. Meng, B. Lorenz, Y. S. Wang, J. Cmaidalka, Y. Y. Xue, and C. W. Chu, Physica C
      353, 195 (2001).
[12]  G. V. M. Williams and S. Krämer, Phys. Rev. B 62, 4132 (2000).
[13]  H. Takagiwa, J. Akimitsu, H. Kawano-Furukawa, and H. Yoshizawa, J. Phys. Soc. Jpn.
      70, 333 (2001).
[14]  E. Takayama-Muromachi, T. Kawashima, N. D. Zhigadlo, T. Drezen, M. Isobe,
      A. T. Matveev, K. Kimoto, and Y. Matsui, Physica C 357-360, 318 (2001).
[15]  T. Kawashima and E. Takayama-Muromachi, Physica C 398, 85 (2003).
[16]  A. C. McLaughlin, W. Zhou, J. P. Attfield, A. N. Fitch, and J. L. Tallon, Phys. Rev B
      60, 7512 (1999).
[17]  J. W. Lynn, B. Keimer, C. Ulrich, C. Bernhard, and J. L. Tallon, Phys. Rev. B. 61,
      R14964, (2000).
[18]  O. Chmaissem, J. D. Jorgensen, H. Shaked, P. Dollar, and J. L. Tallon, Phys. Rev. B 61,
      6401 (2000).
[19]  J. D. Jorgensen, O. Chmaissem, H. Shaked, S. Short, P. W. Klamut, B. Dabrowski,
      and J. L. Tallon, Phys. Rev. B 63, 54440 (2001).
[20]  A. Butera, A. Fainstein, E. Winkler, and J. Tallon, Phys. Rev. B 63, 54442 (2001).

[21]   R. S. Liu, L. -Y. Jang, H. -H. Hung, and J. L. Tallon, Phys. Rev. B $\underline{63}$, 212507 (2001).
[22]   K. Kumagai, S. Takada, and Y. Furukawa, Phys. Rev. B $\underline{63}$, 180509 (2001).
[23]   Y. Tokunaga, H. Kotegawa, K. Ishida, Y. Kitaoka, H. Takagiwa, and J. Akimitsu, Phys. Rev. Lett. $\underline{86}$, 5767 (2001).
[24]   Y. Furukawa, S. Takada, K. Kumagai, T. Kawashima, E. Takayama-Muromachi, N. Kobayashi, T. Fukase, K. Chiba, and T. Goto, J. Phys. Chem. Solids $\underline{63}$, 2315 (2002).
[25]   H. Sakai, N. Osawa, K. Yoshimura, M. Fang, and K. Kosuge, Phys. Rev. B $\underline{67}$, 184409 (2003).
[26]   J. L. Tallon, J. W. Loram, G. V. M. Williams, and C. Bernhard, Phys. Rev. B $\underline{61}$, R6471 (2000).
[27]   I. Felner, U. Asaf, S. Reich, and Y. Tsabba, Physica C $\underline{311}$, 163 (1999).
[28]   P. W. Klamut, B. Dabrowski, S. M. Mini, M. Maxwell, J. Mais, I. Felner, U. Asaf, F. Ritter, A. Shengelaya, R. Khasanov, I. M. Savic, H. Keller, A. Wisniewski, R. Puzniak, I. M. Fita, C. Sulkowski, and M. Matusiak, Physica C $\underline{387}$, 33 (2003).
[29]   G. R. Blake, P. G. Radelli, J. D. Jorgensen, P. W. Klamut, B. Dabrowski, and O. Chmaissem, European Conf. Neutron Scattering, 2003, to be published.
[30]   P. W. Klamut, B. Dabrowski, S. Kolesnik, M. Maxwell, and J. Mais, Phys. Rev. B $\underline{63}$, 224512 (2001).
[31]   C. Bernhard, J. L. Tallon, E. Brücher, and R. K. Kramer, Phys. Rev. B $\underline{61}$, R14960 (2000).
[32]   V. P. S. Awana, S. Ichihara, J. Nakamura, M. Karppinen, H. Yamauchi, J. Yang, W. B. Yelon, W. J. James, and S. K. Malik, J. Appl. Phys. $\underline{91}$, 8501 (2002).
[33]   T. Yokosawa, V. P. S. Awana, K, Kimoto, E. Takayama-Muromachi, M. Karppinen, H. Yamauchi, and Y. Matsui, Ultramicroscopy, to be published.
[34]   N. D. Zhigadlo, P. Odier, J. Ch. Marty, P. Bordet, and A. Sulpice, Physica C $\underline{387}$, 347 (2003).
[35]   Y. Y. Xue, F. Chen, J. Cmaidalka, R. L. Meng, and C. W. Chu, Phys. Rev. B $\underline{67}$, 224511 (2003).
[36]   B. Lorenz, Y. Y. Xue, and C. W. Chu, <u>Studies of High-Temperature Superconductors</u>, Vol. 46, ed. A. V. Narlikar (Nova Science, New York, 2004).
[37]   P. W. Klamut, B. Dabrowski, S. M. Mini, M. Maxwell, S. Kolesnik, J. Mais, A. Shengelaya, R. Khasanov, I. Savic, H. Keller, T. Graber, J. Gebhardt, P. J. Viccaro, and Y. Xiao, Physica C $\underline{364-365}$, 313 (2001).
[38]   P.W. Klamut, B. Dabrowski, M. Maxwell, J. Mais, O. Chmaissem, R. Kruk, R. Kmiec, and C.W. Kimball, Physica C $\underline{341-348}$, 455 (2000)
[39]   W. G. Clark, private communication.
[40]   Y. D. Zhang, J. I. Budnick, J. C. Ford, and W. A. Hines, J. Magn. Magn. Mat. $\underline{100}$, 13 (1991) and references therein.
[41]   J. R. O'Brien, private communication.
[42]   T. P. Papageorgiou, H. F. Braun, and T. Herrmannsdörfer, Phys. Rev. B $\underline{66}$, 104509 (2002).
[43]   Magnetic Resonance Table, http://bmrl.med.uiuc:8080mritable/
[44]   R. Kruk, R. Kmiec, P. W. Klamut, B. Dabrowski, D. E. Brown, M. Maxwell, and C. W. Kimball, Physica C $\underline{370}$, 71 (2002).
[45]   M. DeMarco, D. Coffey, J. Tallon, M, Haka, S. Toorongian, and J Fridman, Phys. Rev. B $\underline{65}$, 212506 (2002).
[46]   H. Mukuda, K. Ishida, Y. Kitaoka, K. Asayama, R. Kanno, and M. Takano, Phys. Rev. B $\underline{60}$, 12279 (1999).

[47] P. D. Battle and W. J. Macklin, J. Solid State Chem. 52, 138 (1984).
[48] B. C. Chakoumakos, S. E. Nagler, S. T. Misture, and H. M. Christen, Physica B 241-243, 358 (1998).
[49] M. Daniel, J. I. Budnick, W. A. Hines, Y. D. Zhang, W. G. Clark, and A. R. Moodenbaugh, J. Phys.: Condens. Matter 12, 3857 (2000)
[50] G. Cao, S. McCall, M. Shepard, J. E. Crow, and R. P. Guertin, Phys. Rev. B 56, 321 (1997).

Table I

| Sample | $T_{mag}$ (K) | $T_{SC\ onset}$ (K) |
|---|---|---|
| (a) Non-SC $RuSr_2EuCu_2O_8$ | 145 | -- |
| (b) Partially-SC $RuSr_2EuCu_2O_8$ | 132 | 28 |
| (c) Non-SC $RuSr_2GdCu_2O_8$ | 138 | -- |
| (d) SC $RuSr_2GdCu_2O_8$ | 130 | 45 |

Table II

| Sample | $T_o$ (K) | $\mu_z$/Ru ($\mu_B$) | $\Theta$ (K) | $\mu_{eff}$/Ru ($\mu_B$) | $\chi_o$ (emu/g) |
|---|---|---|---|---|---|
| (a) Non-SC $RuSr_2EuCu_2O_8$ | 145 | 0.99 | 132 | $3.2_0$ | $1.3 \times 10^{-5}$ |
| (b) Partially-SC $RuSr_2EuCu_2O_8$ | 132 | 0.95 | 118 | $3.1_9$ | $1.6 \times 10^{-5}$ |

Table III

| Sample | Ru$^{4+}$ | | | Ru$^{5+}$ | | |
|---|---|---|---|---|---|---|
| | $H_{hf}$ (kOe) | $\nu_Q$ (MHz) $^{99}$Ru | $\nu_Q$ (MHz) $^{101}$Ru | $H_{hf}$ (kOe) | $\nu_Q$ (MHz) $^{99}$Ru | $\nu_Q$ (MHz) $^{101}$Ru |
| (a) Non-SC RuSr$_2$EuCu$_2$O$_8$ | -366 | -- | -- | -586.6 | 2.9 | 17.0 |
| (b) Partially-SC RuSr$_2$EuCu$_2$O$_8$ | -290 | -- | -- | -587.7 | 2.8 | 16.6 |
| (c) Non-SC RuSr$_2$GdCu$_2$O$_8$ | -290 | -- | -- | -586.6 | 2.8 | 16.4 |
| (d) SC RuSr$_2$GdCu$_2$O$_8$ | -270 | -- | -- | -587.7 | 2.8 | 16.4 |

FIGURE CAPTIONS

Fig.1. Powder pattern Cu K$\alpha_I$ x-ray diffraction scans ($\lambda$ = 1.5406Å) for the Ru1212 samples studied in this work: (a) non-superconducting RuSr$_2$EuCu$_2$O$_8$, (b) partially-superconducting RuSr$_2$EuCu$_2$O$_8$, (c) non-superconducting RuSr$_2$GdCu$_2$O$_8$, and (d) superconducting RuSr$_2$GdCu$_2$O$_8$. The Miller indices are referred to a tetragonal unit cell (P4/mmm). The peaks (*) at 32.2º, 46.2º, and 57.6º for the (c) partially-superconducting RuSr$_2$EuCu$_2$O$_8$ sample are the (200), (220), and (312) reflections for the SrRuO$_3$ impurity phase. The impurity phase is not seen in the patterns for the other samples. The peaks (marked as "base") at 43.5º and 50.6º are background from the sample holder and not associated with the samples.

Fig.2. Zero-field-cooled and field-cooled magnetization versus temperature at the indicated magnetic field for the Ru1212 samples studied in this work: (a) non-superconducting RuSr$_2$EuCu$_2$O$_8$, (b) partially-superconducting RuSr$_2$EuCu$_2$O$_8$, (c) non-superconducting RuSr$_2$GdCu$_2$O$_8$, and (d) superconducting RuSr$_2$GdCu$_2$O$_8$. All samples show magnetic ordering at a temperature between 135 K and 145 K. Both magnetization curves for sample (d), superconducting RuSr$_2$GdCu$_2$O$_8$, show a peak-like anomaly upon entering the superconducting state (see text).

Fig.3. Temperature dependencies of the real part of the *ac* susceptibility (H$_{ac}$=1 Oe, f=200 Hz) for RuSr$_2$GdCu$_2$O$_8$ batch (b) - open circles, and RuSr$_2$EuCu$_2$O$_8$ batch (d) - closed circles. T$_{c1}$ mark the onset of the superconducting transition, T$_{c2}$ - the onset temperature for bulk screening currents in the samples.

Fig.4. Full hysteresis loops measured at 5.0 K: (a) non-superconducting $RuSr_2EuCu_2O_8$ and (b) partially-superconducting $RuSr_2EuCu_2O_8$. Although the magnetization does not saturate in either case, Ru moment values of 0.99 $\mu_B$ and 0.95 $\mu_B$ were obtained at 50 kG for (a) non-superconducting $RuSr_2EuCu_2O_8$ and (b) partially-superconducting $RuSr_2EuCu_2O_8$, respectively, in the magnetically-ordered state.

Fig.5. Curie-Weiss analysis for: (a) non-superconducting $RuSr_2EuCu_2O_8$ and (b) partially-superconducting $RuSr_2EuCu_2O_8$. Ru moment values of $3.2_0$ $\mu_B$ and $3.1_9$ $\mu_B$ were obtained for (a) non-superconducting $RuSr_2EuCu_2O_8$ and (b) partially-superconducting $RuSr_2EuCu_2O_8$, respectively, in the paramagnetic state (see Table II).

Fig.6. Zero-field spin-echo $^{99,101}$Ru NMR and $^{63,65}$Cu NQR spectra at 1.3 K for the Ru1212 samples studied in this work: (a) non-superconducting $RuSr_2EuCu_2O_8$, (b) partially-superconducting $RuSr_2EuCu_2O_8$, (c) non-superconducting $RuSr_2GdCu_2O_8$, and (d) superconducting $RuSr_2GdCu_2O_8$. The high frequency spectra (110 to 150 MHz), characterized by distinct peaks, are attributed to $^{99,101}$Ru NMR signals arising from Ru in the $Ru^{5+}$ valence state. The low frequency spectra (40 to 90 MHz), characterized by considerable broadening and quadrupole features that are barely distinguishable, are attributed to $^{99,101}$Ru NMR signals arising from Ru in the $Ru^{4+}$ valence state. Spectral features attributed to $^{63,65}$Cu NQR with considerable broadening are also present (26 to 34 MHz).

Fig.7. Zero-field spin-echo $^{101}$Ru NMR central (-1/2 $\leftrightarrow$ +1/2) peak at 1.3 K for $Ru^{5+}$ with various applied magnetic fields. Both (a) and (b) above are for the superconducting $RuSr_2GdCu_2O_8$ sample. Similar behavior was observed for the partially-superconducting $RuSr_2EuCu_2O_8$ sample. The inset in Fig. 7a shows that the entire five peak spectrum is shifted with no change in the splitting.

Fig.8. From Fig. 7, central peak height (amplitude maximum) as a function of the applied magnetic field: (a) partially-superconducting $RuSr_2EuCu_2O_8$ and (b) superconducting $RuSr_2GdCu_2O_8$. For both cases, the peak height increases, reaching a maximum between 4 to 6 kOe, and then decreases. This behavior is not typical for a bulk ferromagnetic material (see text).

Fig.9. From Fig. 7, central peak position as a function of the applied magnetic field: (a) partially-superconducting $RuSr_2EuCu_2O_8$ and (b) superconducting $RuSr_2GdCu_2O_8$. For both cases, the peak position shifts to lower frequency with a linear behavior for higher fields. This behavior is not typical for a bulk polycrystalline antiferromagnetic material (see text).

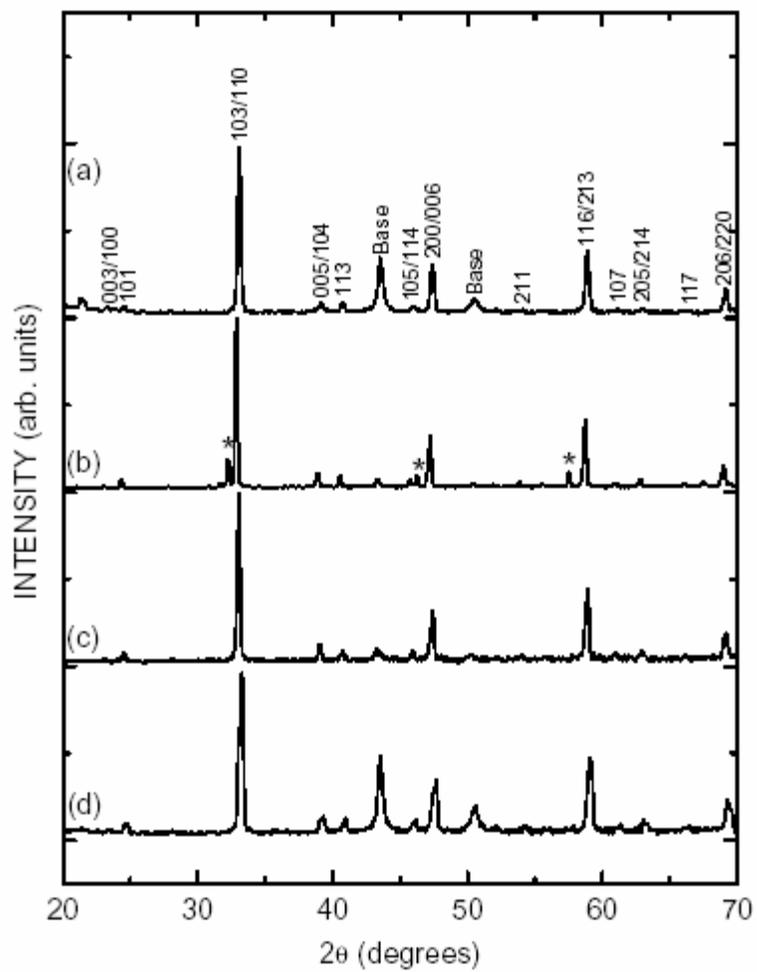

Fig.1    Z.H. Han et al.

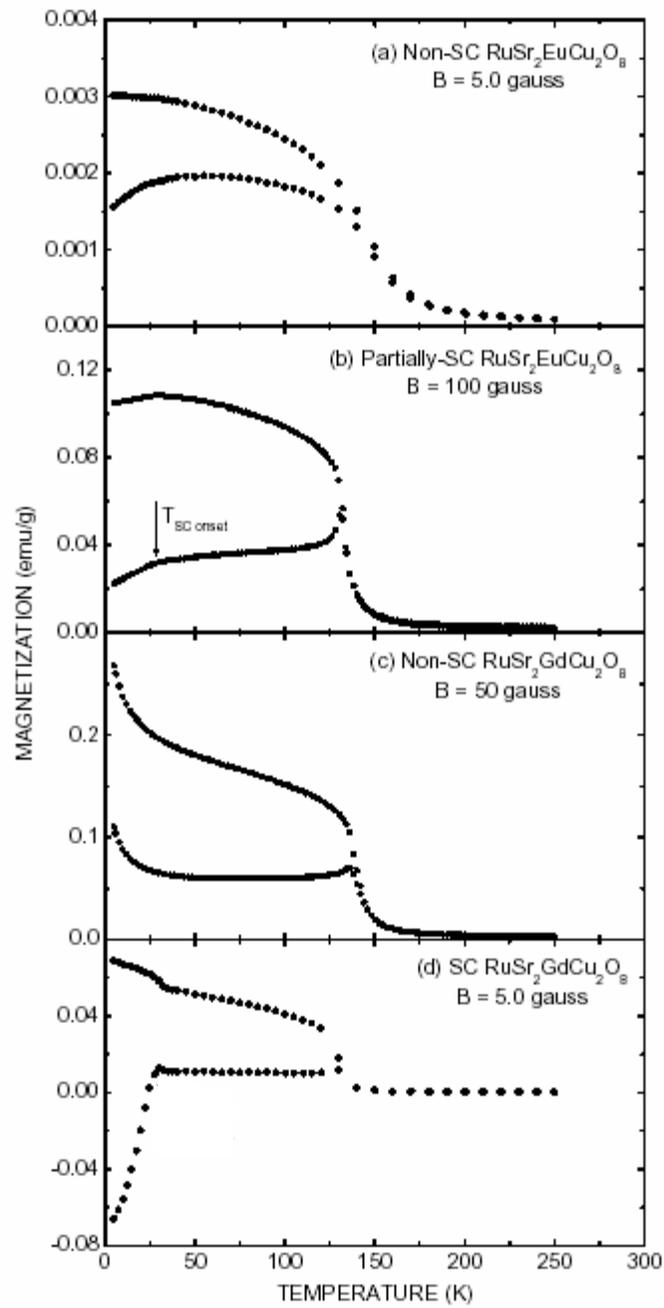

Fig.2    Z.H. Han et al.

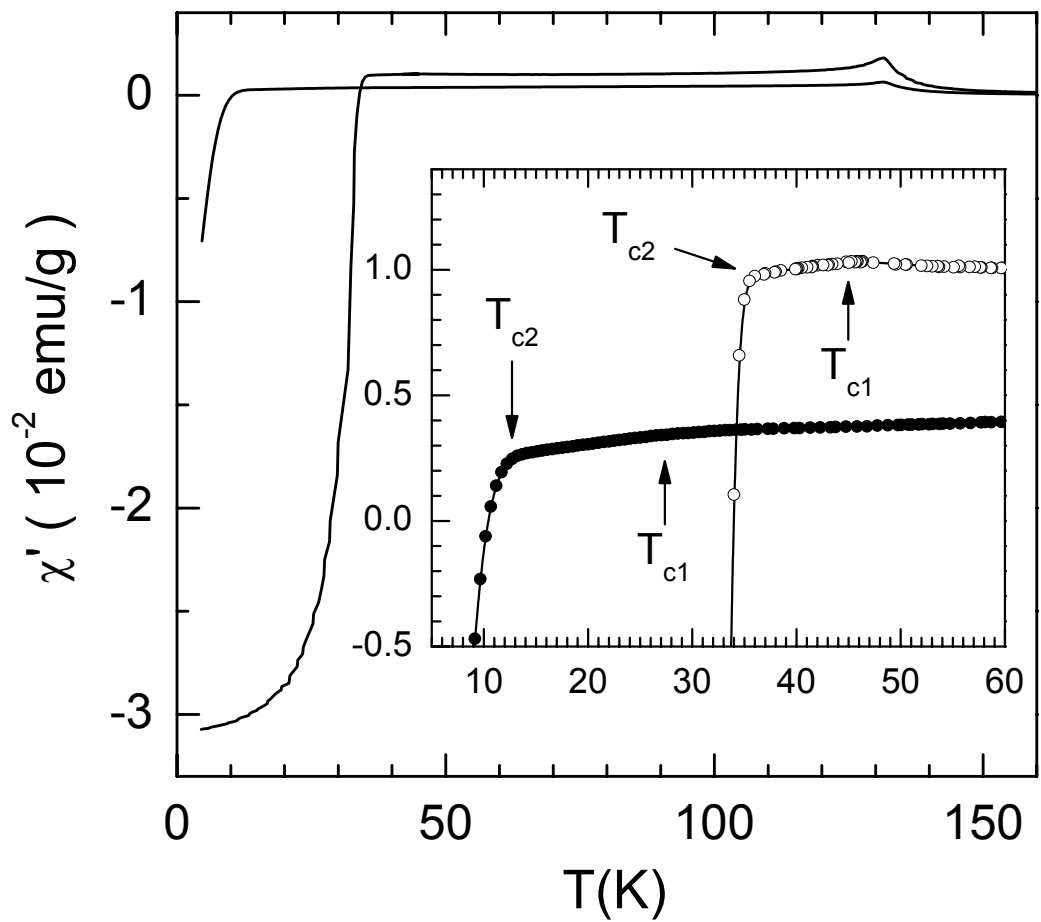

Fig.3    Z.H. Han et al.

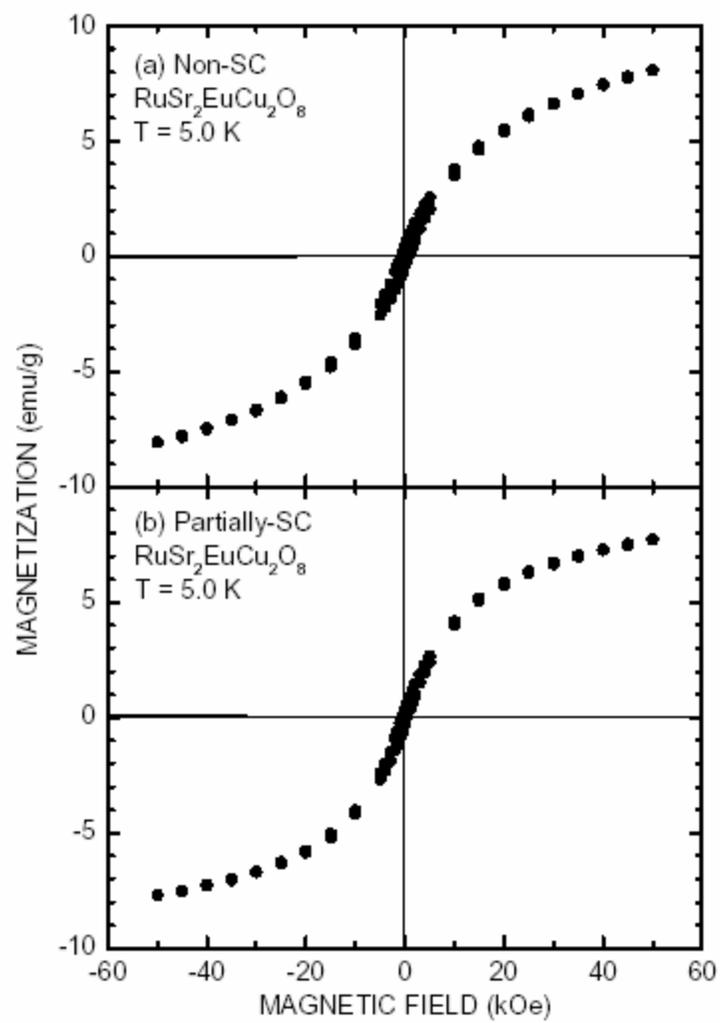

Fig.4    Z.H. Han et al.

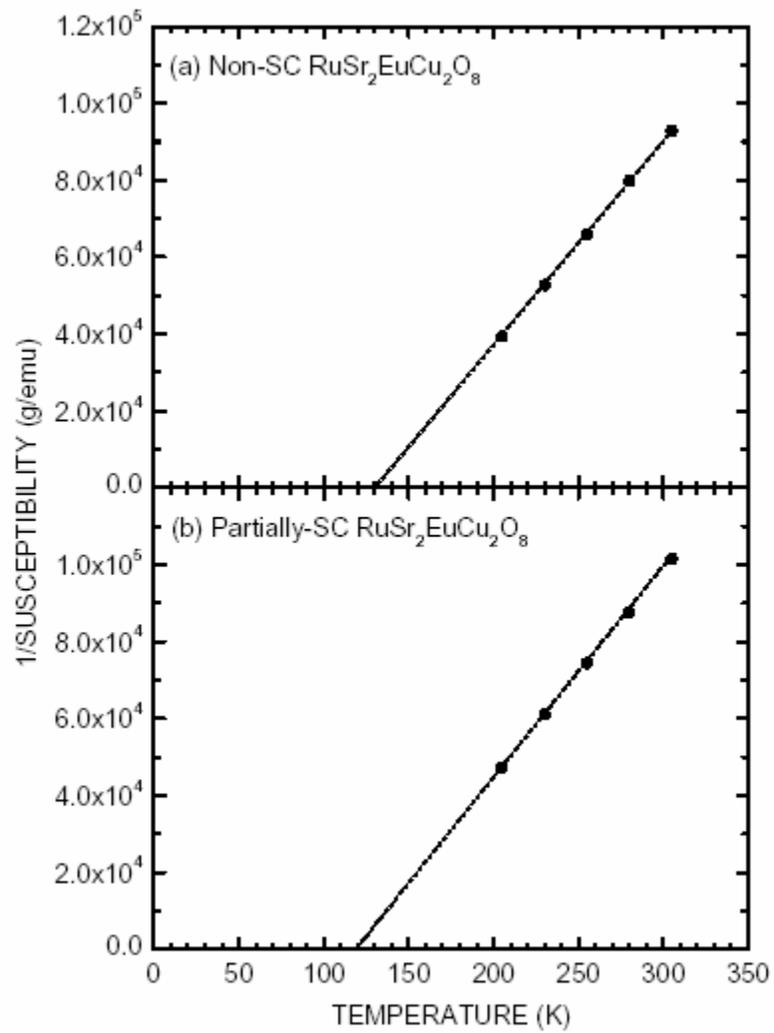

Fig.5    Z.H. Han et al.

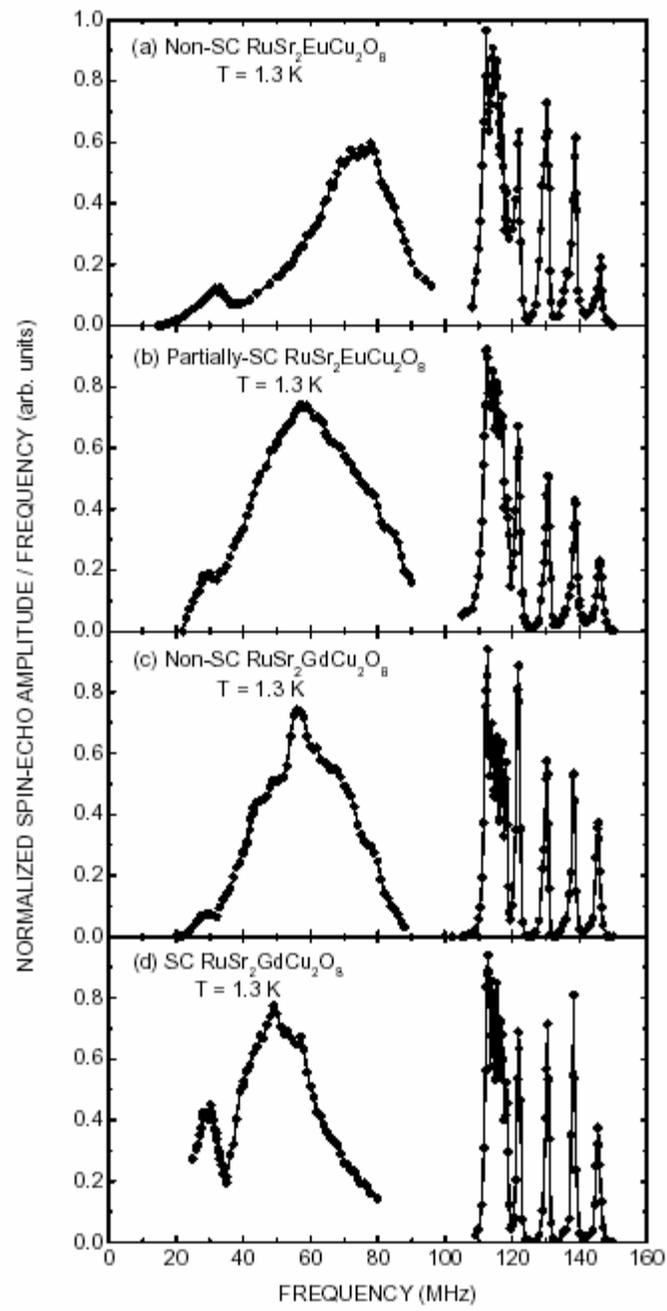

Fig.6    Z.H. Han et al.

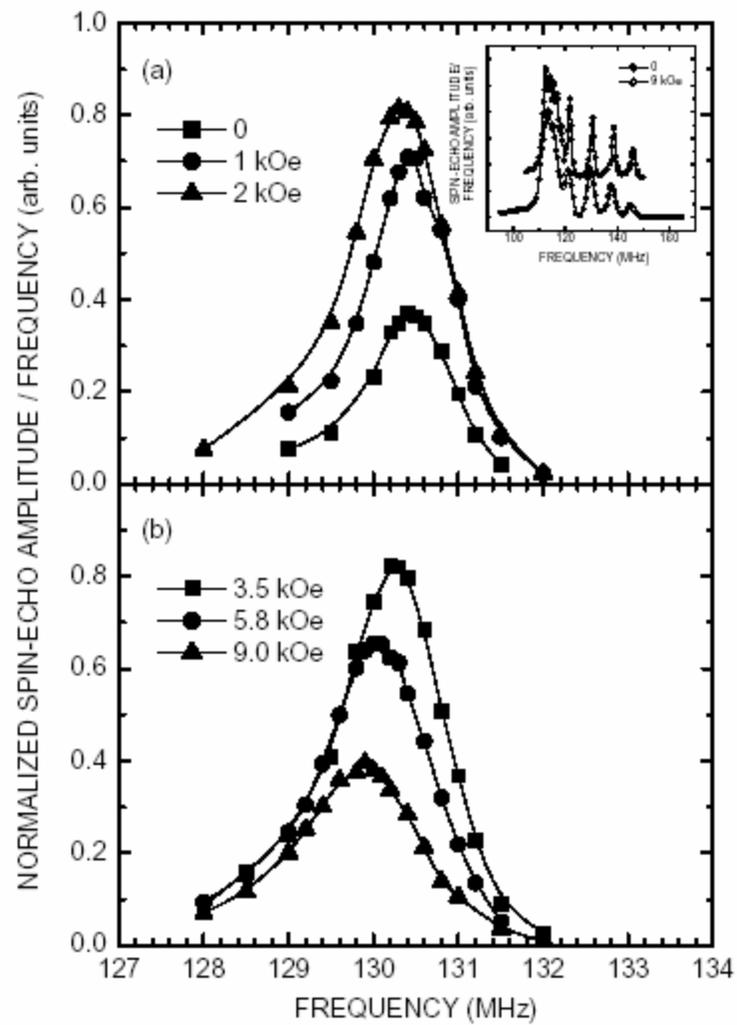

Fig.7    Z.H. Han et al.

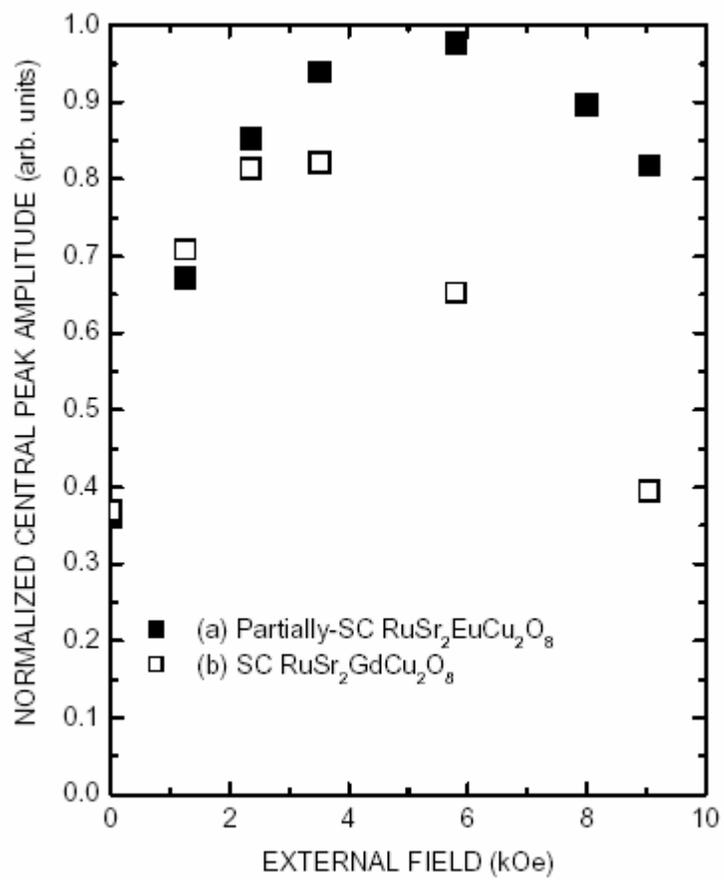

**Fig.8    Z.H. Han et al.**

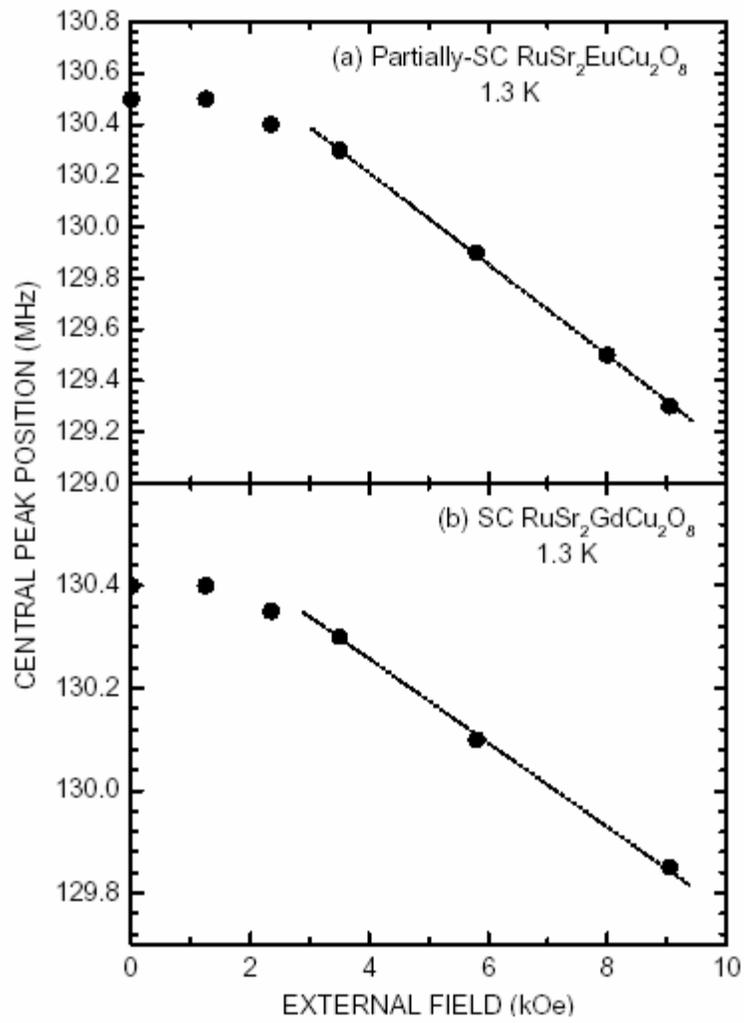

**Fig.9    Z.H. Han et al.**